\documentclass[doublecol]{epl2} 

\usepackage{amsmath,amssymb,slashbox,color}

\newcommand{\io}
   {\mathrel{\rlap{\raise1.5pt\hbox{$<$}}{\lower3pt\hbox{$\sim$}}}}

\newcommand{\so}
   {\mathrel{\rlap{\raise1.5pt\hbox{$>$}}{\lower3pt\hbox{$\sim$}}}}
   
\definecolor{gris}{gray}{0.66}

\title{Universal amplitudes of the Casimir-like interactions between four types of rods in fluid membranes}
\shorttitle{Casimir interactions between various types of rods} 

\author{Anne-Florence Bitbol, K\'evin Sin Ronia \and Jean-Baptiste Fournier}
\shortauthor{A.-F. Bitbol \etal}

\institute{                    
  Universit\'e Paris Diderot, Sorbonne Paris Cit\'e, Laboratoire Mati\`ere et Syst\`emes Complexes (MSC), UMR 7057 CNRS, F-75205 Paris, France, EU}
  

\abstract{
The fluctuation-induced, Casimir-like interaction between two parallel rods of length $L$ adsorbed on a fluid membrane is calculated analytically at short separations $d\ll L$. The rods are modeled as constraints imposed on the membrane curvature along a straight line. This allows to define four types of rods, according to whether the membrane can twist along the rod and/or curve across it. For stiff constraints, all the interaction potentials between the different types of rods are attractive and proportional to $L/d$.
Two of the four types of rods are then equivalent, which yields six universal Casimir amplitudes. Repulsion can occur between different rods for soft constraints. Numerical results obtained for all ranges of $d/L$ show that the attraction potential reaches $k_\mathrm{B}T$ for $d/L\simeq0.2$. At separations smaller than $d_c\approx L(L/\ell_p)^{1/3}$, where $\ell_p$ is the rod persistence length, two rods with fixed ends will bend toward each other and finally come into contact because of the Casimir interaction.
}

\begin{document}

\maketitle

When a field with long-range correlations is confined between objects that constrain its fluctuations, generic long-range forces arise. The most famous example is the Casimir interaction between two uncharged metal plates, which originates from the constraints they impose on the quantum electrodynamical vacuum fluctuations~\cite{Casimir48}. Similar interactions arise between objects immersed in a classical fluid, due to reduced thermal fluctuations. Fluids with long-range correlations produce long-range Casimir-like forces with universal characteristics. These interactions have been predicted to occur in binary mixtures close to a critical point, superfluids, liquid crystals and fluid membranes or interfaces~\cite{Kardar99,Gambassi09}. Direct measurements of thermal Casimir-like forces have been achieved recently between a colloid and a surface in a critical binary mixture~\cite{Hertlein08}. 

In biomimetic membranes, which are fluid interfaces characterized by their bending rigidity~\cite{Helfrich73}, Casimir-like forces have been predicted to arise between phase-separated domains~\cite{Goulian93}, integral proteins~\cite{Goulian93}, adsorbed stiff rods~\cite{Golestanian96pre} and semi-flexible polymers~\cite{Golestanian96epl}. All these inclusions impose constraints on the thermal fluctuations of the membrane shape. The interactions between inclusions within biomembranes, and their organization, plays a central role in many cellular regulatory processes and is of fundamental importance~\cite{Alberts_book}. Biologically relevant stiff polymers adsorbed on membranes include actin filaments~\cite{Limozin03,Hase06}, FtsZ rings~\cite{Osawa08} and the $\beta$-Amyloid fibrils involved in Alzheimer's disease~\cite{Bokvist04, Williams11}. Adsorption of goethite ($\alpha$--FeOOH) nanorods on membranes has also been reported~\cite{Constantin10}.

In this paper, we study the fluctuation-induced Casimir-like interaction free energy $\mathcal{F}_\ab{C}$ between two parallel rods of length $L$ adsorbed on a membrane\footnote{Our model is valid both for embedded and for adsorbed rods.}, at short and intermediate separations $d\io L$. It can be calculated by studying rods whose only effect is to restrict the membrane fluctuations: then, $\mathcal{F}_\ab{C}$ corresponds to the $d$-dependent part of the free energy of the system. The Casimir-like force can be obtained by direct differentiation of $\mathcal{F}_\ab{C}$. We focus on membranes dominated by their bending rigidity $\kappa$, i.e., we assume that the length scale $(\kappa/\sigma)^{1/2}$, which compares its effect to that of the membrane tension $\sigma$, is much larger than all other relevant lengths. Other cases are briefly discussed at the end of this paper. In the asymptotic limit $d\gg L$, previous analytical calculations showed that $\mathcal{F}_\ab{C}$ is attractive and decays as $k_\ab{B}T(L/d)^4$~\cite{Golestanian96pre}, where $k_\ab{B}$ is Botzmann's constant, and $T$ is the temperature. Thus, even for $d\simeq3L$, its magnitude $\approx 10^{-4}k_\ab{B}T$ is far too small to effectively bring rods together. At short separations $d\ll L$, one expects a constant scale-free Casimir interaction per unit length, i.e., $\mathcal{F}_\ab{C}\propto-k_\ab{B}T L/d$. Effective attraction at $d\io L$ will thus result if the numerical prefactor is of order unity. Golestanian~\cite{Golestanian96epl} showed that the finite bending rigidity of semiflexible polymers, or polymer rods, embedded in membranes, has dramatic effects on their Casimir interaction at short separations: shape fluctuations, which locally reduce the separation of the polymer rods, become unstable. Precisely, the free energy eigenvalues of these modes were shown to be $\lambda(q)\simeq(\kappa_r/k_\ab{B}T)q^4-(2A/\pi)/d^3$, where $\kappa_r$ is the bending rigidity of the polymer rods and $A\simeq2.91$. Instabilities occur if $\lambda(q_\ab{min})<0$ with $q_\ab{min}\approx\pi/L$, i.e., for $d<d_c$, where
\begin{equation}
\label{dc}
\frac{d_c}{L}\approx\left(\frac{2A}{\pi^5}\right)^{1/3}\left(\frac{L}{\ell_p}\right)^{1/3}.
\end{equation}
Here, $\ell_p=\kappa_r/(k_\ab{B}T)$ is the persistence length of the rod. For actin filaments~\cite{Limozin03,Hase06} of length $L\simeq100\un{nm}$, with $\ell_p\simeq17\un{\mu m}$, this yields $d_c/L\approx5\times10^{-2}$. For $L/\ell_p\simeq10^{-4}$, which may be realistic for actin bundles, the undulation instability threshold goes down to $d_c/L\approx10^{-2}$. It is thus realistic to consider stiff rods at separations in the range $0.01\io d/L\io 1$. In this paper we shall give analytical results for $d\ll L$, and numerical results for all ranges of~$d/L$.

\begin{figure}
\onefigure[scale=.35]{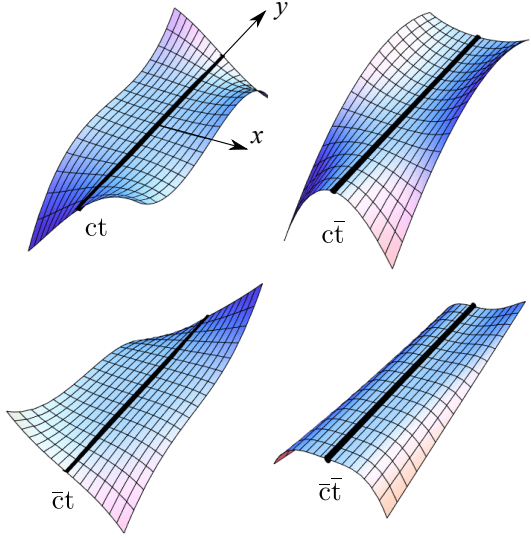}
\caption{Four types of rods for which the membrane may or may not curve (c) across the rod or twist (t) along it.}
\label{f.rods}
\end{figure}

Let us describe the shape of the membrane by its height $z=h(\vect{r})$ above a reference plane parametrized by $\vect{r}=(x,y)$. We shall refer to the latter coordinates as ``in-plane" and to $z$ as ``out-of-plane". In the Gaussian limit of small deformations around the flat shape, the effective Hamiltonian of the membrane reads~\cite{Helfrich73}:
\begin{equation}
\mathcal{H}=\int\!\upd\vect{r}\,\frac{\kappa}{2}\left(\nabla^2h\right)^2.
\end{equation}
Note that the Gaussian curvature does not contribute, due to the Gauss-Bonnet theorem~\cite{Chaikin_book}, and that the membrane tension $\sigma$ has been neglected as previously discussed. The presence of a rigid rod lying within the membrane (or absorbed on it) is traditionally described by imposing the constraint~\cite{Goulian93,Golestanian96pre}:
\begin{equation}
h(\vect{r})=a+\vect{b}\cdot\vect{r}
\end{equation}
for $\bm{r}$ in a narrow rectangle $[x_0-\epsilon,x_0+\epsilon]\times[0,L]$, where $a$ and $\vect{b}$ describe translational and tilt degrees of freedom, respectively. This is fully equivalent to imposing in the same domain the constraints:
\begin{equation}
h_{yy}=0\,,\qquad h_{xx}=0\,,\qquad h_{xy}=0\,,
\label{cond}
\end{equation}
where $h_{ij}\equiv\partial_i\partial_jh$. An infinitely narrow rod can be described by imposing conditions~(\ref{cond}) for $x=x_0$ and $y\in[0,L]$. The single condition $h_{yy}=0$, however, is sufficient to specify a rod. In the absence of the other conditions, the membrane can curve across the rod and twist along it. It follows that we may consider four types of rods (fig.~\ref{f.rods}):
\begin{enumerate}
\item Curved-twisted ($\mathrm{ct}$): $h_{yy}=0$,
\item Curved-nontwisted ($\mathrm{c\bar t}$): $h_{yy}=h_{xy}=0$,
\item Noncurved-twisted ($\mathrm{\bar ct}$): $h_{yy}=h_{xx}=0$,
\item Noncurved-nontwisted ($\mathrm{\bar c\bar t}$): $h_{yy}=h_{xx}=h_{xy}=0$.
\end{enumerate}
The actual constraints imposed on the coarse-grained field $h$ by a given rod depend on its microscopic properties, such as its width and its interactions with the lipids that constitute the membrane. Let us introduce the natural short length cutoff $a$ of the theory, which is of order of the membrane thickness (it cannot be smaller since the membrane is considered as a thin surface): the field $h$ is constructed through a coarse-graining, for instance a simple moving average, of the height of each lipid molecule over the lengthscale $a$. First, let us consider a rod with width smaller than $a$ that interacts with the lipids through short-range (e.g. hydrophobic) interactions, which can be neglected at a distance $a$. Such a rod cannot affect $h_{xx}$ or $h_{xy}$, and it will thus be of the ``ct" type. On the contrary, a rod with a width equal to, or a little bit larger than $a$ will be of the ``$\mathrm{\bar c \bar t}$" type. Indeed, this width is sufficient for the thin flat region arising from the presence of the rod to remain after the coarse-graining. A rod which has strong anisotropic attractive interactions with the lipids, with a range at least of order $a$, would behave as a ``$\mathrm{\bar c \bar t}$" rod too. In the case of intermediate ranges of interactions, the rod could constrain the slope but not the curvature of the membrane, in which case one would obtain the ``$\mathrm{c\bar t}$" type. Finally, the ``$\mathrm{\bar ct}$" type could correspond to a rod with width of order $a$ and with short-range interactions, but composed of units free to twist relative to one another. In our coarse-grained description, all such rods can be considered as one-dimensional since their width is of order $a$ or smaller. Hence, the microscopic properties of a rod (width, interactions) will appear only indirectly in our work, through the rod type.

Let us first calculate the Casimir interaction between two parallel rods of the simplest ``ct" type, in the $d\ll L$ regime. Let us assume periodic boundary conditions with period $L$ along the $y$ direction. Generalizing the constraint $h_{yy}=0$ to rods with a finite rigidity, we account for the presence of the rods by adding to $\mathcal{H}$ the following interaction energies between each rod and the membrane:
\begin{eqnarray}
\label{finite}
\mathcal{H}'_i=\int_0^L\!\upd y\,\frac{\kappa_\perp}{2}h_{yy}^2(x_i,y),
\end{eqnarray}
for $i\in\{1,2\}$, with $x_1=0$ and $x_2=d>0$. Thus, we allow out-of-plane fluctuations of the rods with a bending rigidity $\kappa_\perp=k_\ab{B}T\ell_\perp$. The hard constraint $h_{yy}=0$ is recovered in the limit $\kappa_\perp\to\infty$. Note that we implicitly assume in any case an infinite in-plane rigidity $\kappa_\parallel\to\infty$, which is justified if $d/L$ is not too small, as discussed above. Adapting the field-theoretical method of Li and Kardar~\cite{Li92} to these general constraints, we introduce two auxiliary fields $\psi_i(y)$, and we make the substitution
\begin{equation}
e^{-\beta\mathcal{H}'_i}=\int\!\mathcal{D}\psi_i\,
e^{-\frac{1}{2}\int\!\upd y\,\ell_\perp^{-1}\psi_i^2
+i\int \upd\vect{r}\,h(\vect{r})\psi_i''(y)\delta(x-x_i)}
\end{equation}
in the partition function $Z=\int\!\mathcal{D}h\,\exp[-\beta(\mathcal{H}+\mathcal{H}'_1+\mathcal{H}'_2)]$ of the membrane containing the two rods, where $\beta=(k_\mathrm{B}T)^{-1}$. Performing the Gaussian integral over $h(\vect{r})$ yields
\begin{equation}
Z=\int\!\mathcal{D}\psi_1\mathcal{D}\psi_2\,
e^{-\frac{1}{2L}\sum_q\Psi^t(-q)M(d,q)\Psi(q)},
\end{equation}
where $\psi_i(q)=\int_0^L\psi_i(y)\exp(-iqy)$, $\Psi(q)=(\psi_1(q)~�\psi_2(q))^t$,
\begin{equation}
M(d,q)=
\begin{pmatrix}
\ell_\perp^{-1}+\Gamma(0,q) & \Gamma(-d,q)\cr \Gamma(d,q) & \ell_\perp^{-1}+\Gamma(0,q)
\end{pmatrix}
\end{equation}
and
\begin{eqnarray}
\Gamma(x_i,q)\!\!&=&\!\!\frac{k_\ab{B}T}{\kappa}q^4\!\int_{-\infty}^\infty\frac{\upd p}{2\pi}
\,\frac{e^{ipx_i}}{\left(p^2+q^2\right)^2}\nonumber\\
&=&\!\!\frac{k_\ab{B}T}{4\kappa}|q|\left(1+|q|x_i\right)e^{-|q|x_i}
\end{eqnarray}
is the Fourier transform along y of $\Gamma(x_i,y)=\partial^4G(x_i,y)/\partial y^4$, where $G(\vect{r}-\vect{r}')=\langle\vect{r}|(\beta\kappa\nabla^4)^{-1}|\vect{r}'\rangle$ is the correlation function of the free membrane. Performing the Gaussian integrals over the fields $\psi_i(q)$ yields the free energy $-k_\ab{B}T\ln Z$ for two parallel ``ct" rods. Discarding $d$-independent terms, we obtain the Casimir interaction
\begin{eqnarray}
\label{integ}
\mathcal{F}_\ab{C}^\mathrm{ct|ct}(d)\!\!&=&\!\!\frac{1}{2}k_\ab{B}T\sum_q\ln\det\left[ M(d,q)\right]\nonumber\\
&=&\!\!\frac{k_\ab{B}TL}{2}\int\!\frac{\upd q}{2\pi}\,\ln\left[
1-\left(\frac{\Gamma(d,q)}{\ell_\perp^{-1}+\Gamma(0,q)}\right)^2
\right]\nonumber\\\label{resctct}
&=&\!\!k_\ab{B}T\frac{L}{d}\,\phi^\mathrm{ct|ct}(d)\,,
\end{eqnarray}
where
\begin{eqnarray}
\label{phi}
\phi^\mathrm{ct|ct}(d)\!\!&=&\!\!\int_0^\infty\!\frac{\upd x}{2\pi}
\ln\left[1-\left(\frac{1+x}{1+\bar d/x}\right)^2e^{-2x}\right],\\
\bar d\!\!&=&\!\!\frac{4\kappa}{k_\ab{B}T}\,\frac{d}{\ell_\perp}\,.
\end{eqnarray}
It can be shown that this integral expression  of the Casimir-like interaction is consistent with eqs.~(4) and (8) of ref.~\cite{Golestanian96epl}. Our method for calculating the Casimir-like interaction, however, is more direct, and we shall extract explicit forms for the asymptotic behaviors and extend it to other types of rods.

\begin{figure}
\onefigure[scale=.37]{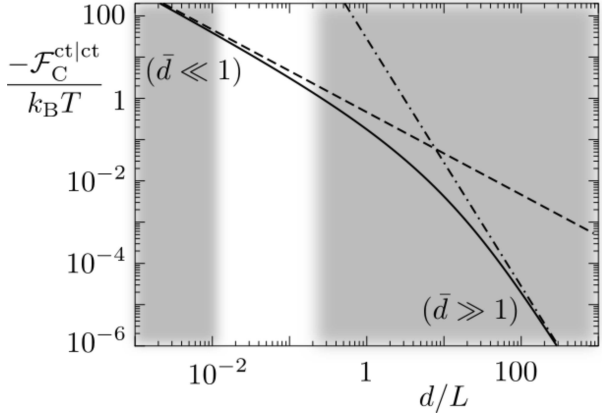}
\caption{Logarithmic plot of the Casimir interaction  (\ref{integ}) between two ``ct" rods of length $L$ and separation $d$, calculated analytically under the assumption $d/L\ll1$, for $L=10^{-3}\ell_p\equiv10^{-3}\ell_\perp$ and $\kappa=25\,k_\mathrm{B}T$ (yielding $\bar d=0.1d/L$). In the right shaded area the condition $d/L\ll1$ is violated, and in the left shaded area the in-plane undulation instability is expected to occur. The dashed lines show the asymptotic regimes (\ref{softregime}) and (\ref{hard}). In the white region, one would measure $\beta\mathcal{F}_\ab{C}^\mathrm{ct|ct}\approx-0.2\,(L/d)^{1.2}$. As discussed in the text, the regime $\bar d\gg1$ could be compatible with $d/L\ll1$ for anisotropic rods with a very soft out-of-plane bending rigidity.}
\label{f.cross}
\end{figure}

Two asymptotic Casimir interaction regimes occur (fig.~\ref{f.cross}). First, in the ``soft" limit where $\bar d$ tends to infinity ($\kappa_\perp\rightarrow 0$), the integral in eq.~(\ref{phi}) can be performed analytically:
\begin{equation}
\label{softregime}
\mathcal{F}_\ab{C,soft}^\mathrm{ct|ct}(d)
\simeq-\frac{7}{8\pi\,\bar d^2}k_\ab{B}T\,\frac{L}{d}=-\frac{7k_\ab{B}T}{128\pi}\,\frac{(\kappa_\perp/\kappa)^2L}{d^3}.
\end{equation}
This interaction, which scales as $d^{-3}$, is quite small, however, since it is proportional to $\bar d^{-2}$. 

Conversely, in the ``hard" limit where $\bar d$ tends to zero ($\kappa_\perp\to\infty$), $\phi^\mathrm{ct|ct}$ reaches a constant limit, denoted by $A^\mathrm{ct|ct}$:
\begin{eqnarray}
\label{hard}
\mathcal{F}_\ab{C}^\mathrm{ct|ct}(d)\!\!&=&\!\!A^\mathrm{ct|ct}\,k_\ab{B}T\frac{L}{d}\,,\\
A^\mathrm{ct|ct}\!\!&=&\!\!\int_0^\infty\!\frac{\upd x}{2\pi}
\ln\left[1-\left(1+x\right)^2e^{-2x}\right]\nonumber\\
&\approx&\!\!-0.46237.
\end{eqnarray}
This ``hard" limit corresponds to rods imposing $h_{yy}=0$. Note that for $\ell_\perp=\beta\kappa_\perp\to\infty$, the $q^4$ factors in $\Gamma(d,q)$ and $\Gamma(0,q)$ cancel each other out in eq.~(\ref{resctct}). Hence, imposing $h_{yy}=0$ on the rod is mathematically equivalent to setting $h=0$ for $x=x_0$ and $y\in[0,L]$ (by means of external forces and torques). This indicates that the relative tilt of the rods is effectively frozen in this regime.
Physically, this comes from the fact that we are in the $d\ll L$ regime: a relative tilt of these long rods would cost a lot of energy. More formally, as we use periodic boundary conditions with period $L$ in the $y$ direction in our calculation, which amounts to neglecting the effect of the edges of the rods, $h_{yy}(x_i,y)=0$ for $y\in[0,L]$ entails that $h(x_i,y)$ is constant. Setting these constants to zero is legitimate since their presence simply corresponds to a possible change of reference plane.

For ordinary rods with $\kappa_r=\kappa_\parallel\approx\kappa_\perp$, which corresponds to $\ell_p=\ell_\parallel\approx\ell_\perp$, it is the ``hard" limit that is the physical one. 
Indeed, $d\ll L$ gives $\bar d\ll4\beta\kappa L/\ell_\perp<1$, since the condition $L<\ell_\perp k_\ab{B}T/(4\kappa)\approx10^{-2}\ell_\perp$ is required to avoid the in-plane undulation instability for small $d/L$, as previously discussed. Hence, the ``hard" interaction, eq.~(\ref{hard}), is the physical one for hard rods at separations $d\ll L$. This asymptotic regime is attained slowly, however, so that one might rather expect to see an intermediate effective power law (see fig.~\ref{f.cross}). In principle, the ``soft" limit can be reached by strongly anisotropic rods with a large $\kappa_\parallel$ and a small $\kappa_\perp$, such that $\ell_\perp k_\ab{B}T/(4\kappa)\ll d\ll L$. Such rods should rather be called ribbons.

Let us now calculate the Casimir interactions between the other pairs of rods in the $d\ll L$ regime. Since the ``hard" limit is more realistic, we shall consider the situation where the conditions $h_{yy}=0$, $h_{xy}=0$ and $h_{xx}=0$ are strictly imposed. Let us define a vectorial differential operator $\mathbb{D}^\alpha$ for each rod type $\alpha\in\{\mathrm{ct},\,\mathrm{c\bar t},\,\mathrm{\bar c t},\,\mathrm{\bar c\bar t}\}$:
\begin{equation}
\begin{tabular}{ll}
$\mathbb{D}^{\mathrm{ct}}=(\partial^2_y)$\,, &
$\mathbb{D}^{\mathrm{c\bar t}}=(\partial_x\partial_y,\,\partial^2_y)^t$\,, \\
$\mathbb{D}^{\mathrm{\bar c t}}=(\partial^2_x,\,\partial^2_y)^t$\,, &
$\mathbb{D}^{\mathrm{\bar c \bar t}}=(\partial^2_x,\,\partial_x\partial_y,\,\partial^2_y)^t$\,.
\end{tabular}
\end{equation}
The constraints imposed by a rod of type $\alpha$ can be written as $\mathbb{D}^\alpha h(x,y)=0$, for $x\in\{0,d\}$ and $y\in[0,L]$. Following the steps detailed above in the case of two ``$\mathrm {ct}$'' rods, we obtain:
\begin{equation}
\mathcal{F}^{\alpha|\beta}_\ab{C}(d)=
\frac{1}{2}k_\ab{B}T\sum_q\ln\det\left[M^{\alpha|\beta}(d,q)\right],
\end{equation}
where $M^{\alpha|\beta}$ is the Hermitian block matrix:
\begin{equation}
M^{\alpha|\beta}(d,q)=
\left(\begin{array}{c|c}
G^{\alpha|\alpha}(0,q) & G^{\alpha|\beta}(-d,q)\\
\hline
G^{\beta|\alpha}(d,q) & G^{\beta|\beta}(0,q)
\end{array}\right)\,,
\label{block}
\end{equation}
in which $(G^{\alpha|\beta})_{ij}=\mathbb{D}^\alpha_i\mathbb{D}^\beta_jG$. 
Note that when $\mathbb{D}^\alpha_i\mathbb{D}^\beta_j$ contains $\partial_x^4$ (e.g., for two ``$\mathrm {\bar ct}$'' rods), the corresponding element of $G^{\alpha|\alpha}(x,q)$ includes a term proportional to $\delta(x)$. This term is discarded in $x=0$, as in Ref.~\cite{Li92}. It is shown further on, in an example, that this regularization procedure is in agreement with zeta-function regularization. We thus obtain the Casimir interaction between a rod of type $\alpha$ and a rod of type $\beta$:
\begin{equation}
\mathcal{F}^{\alpha|\beta}_\ab{C}(d)
=A^{\alpha|\beta}k_\mathrm{B}T\frac{L}{d}\,,
\label{F2}
\end{equation}
with $A^{\alpha|\beta}(=A^{\beta|\alpha})$ given by
\begin{equation}
A^{\alpha|\beta}=\int_0^\infty \frac{\upd x}{2\pi} \ln\left[1-f^{\alpha|\beta}(x)\right]\,.
\label{Aij}
\end{equation}
The function $f^{\alpha|\beta}$ depends on the types of rods at stake:
\begin{eqnarray}
f^{\mathrm{ct|ct}}(x)&=&\left(1+x\right)^2e^{-2x}\,,\\
f^{\mathrm{ct|c\bar t}}(x)&=&\left(1+2\,x+2\,x^2\right)e^{-2x}\,,\\
f^{\mathrm{ct|\bar ct}}(x)&=&\left(1+2\,x\right)e^{-2x}\,,\\
f^{\mathrm{c\bar t|c\bar t}}(x)&=&2\left(1+2\,x^2\right)e^{-2x}-e^{-4x}\,,\\
f^{\mathrm{c\bar t|\bar ct}}(x)&=&4\,x\,e^{-2x}+e^{-4x}\,,\\
f^{\mathrm{\bar ct|\bar ct}}(x)&=&2\,e^{-2x}-e^{-4x}\,.
\end{eqnarray}
We find that the results involving ``$\mathrm{\bar c\bar t}$'' rods are strictly identical to those involving ``$\mathrm{ct}$'' rods: $f^{\mathrm{\bar c \bar t}|\beta}\equiv f^{\mathrm{ct}|\beta}$ for all $\beta$.

\begin{table}
\caption{Universal amplitudes $A^{\alpha|\beta}$ of the Casimir interactions between the various types of rods, rounded off to the third decimal place. The exact value of  $A^{\mathrm{\bar ct|\bar ct}}$ is $-\pi/12$.
}
\label{coefs}
\begin{center}
\begin{tabular}{|c|ccc|}
\hline
\backslashbox{$\alpha$}{$\beta$} & $\mathrm{\bar ct}$ & $\mathrm{ct\,/\,\bar c\bar t}$ & $\mathrm{c\bar t}$\\
\hline
$\mathrm{\bar ct}$ & -0.262 & -0.357 & -0.549 \\
$\mathrm{ct\,/\,\bar c\bar t}$ & & -0.462 & -0.672  \\
$\mathrm{c\bar t}$ & & & -0.924 \\
\hline
\end{tabular}
\end{center}
\end{table}

\begin{figure}
\onefigure[scale=.37]{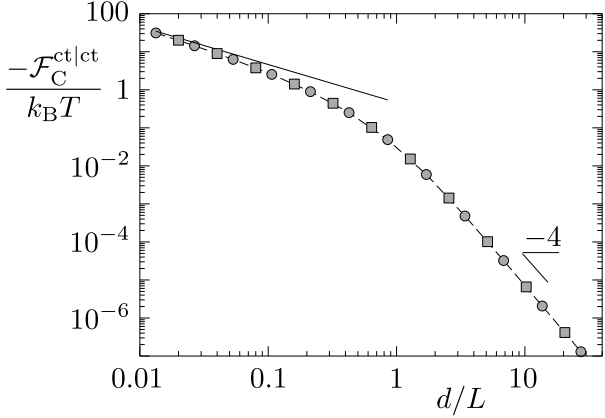}
\caption{Logarithmic plot of the universal Casimir interaction between two rigid ``ct" rods of length $L$ and separation $d$, calculated numerically in all regimes of $d/L$. Circles (resp.\ squares) corresponds to a length $L$ of the rods $\simeq75$ (resp.\ 50) times the microscopic cutoff. The straight line shows the analytical asymptotic behavior $\mathcal{F}^{\mathrm{ct|ct}}_\ab{C}/(k_\mathrm{B}T)=A^\mathrm{ct|ct}\,d/L$, valid for $d/L\ll 1$.}
\label{f.discrete}
\end{figure}

The numerical values of the $A^{\alpha|\beta}$ coefficients are given in table~\ref{coefs}. They are all negative, which means that all the rods attract each other. Casimir interactions are ordinarily attractive for like boundary conditions and repulsive for unlike boundary conditions~\cite{Gambassi09}. Here, however, the situation is not so simple, since all the rods share the $h_{yy}=0$ condition, while they differ on the other ones. We obtain increasing interactions upon going from ``$\mathrm{\bar ct}$'' to ``$\mathrm{ct}$/$\mathrm{\bar c\bar t}$'' then to ``$\mathrm{c\bar t}$''. Besides, as mentioned previously, ``$\mathrm{ct}$'' and ``$\mathrm{\bar c\bar t}$'' rods behave equivalently. However, this equivalence holds only in the ``hard'' limit. Indeed, if we put a finite rigidity $\kappa_\perp$ on $h_{yy}$, as in eq.~(\ref{finite}), the ``$\mathrm{ct}$'' rod vanishes altogether in the limit $\kappa_\perp\to0$ while the ``$\mathrm{\bar c\bar t}$'' rod becomes a different object: a flexible ``ribbon" setting only $h_{xx}=0$. We have calculated the Casimir interaction between this type of ribbon and a ``$\mathrm{ct}$" rod imposing $h_{yy}=0$. We find $A^{\mathrm{rib}|\mathrm{ct}}\simeq0.012>0$, i.e., these antagonistic boundary conditions ($h_{yy}=0$ vs. $h_{xx}=0$) indeed yield a repulsion.

Exact calculations of the fluctuation-induced interactions between rods at short separations $d\ll L$ were made possible by neglecting boundary effects, which results in translational invariance. Formally, this was implemented through periodic boundary conditions. At larger separations, the effect of the edges of the rods cannot be neglected anymore. However, the Casimir interaction energy can be computed numerically for all $d/L$. Numerical schemes based on the Li and Kardar method have been developed in Refs.~\cite{Buscher04,Noruzifar09}. The idea is to discretize the functional integral over the auxiliary fields $\psi_i$ that enforce the constraints. Using a discretization scheme that will be discussed elsewhere~\cite{SinRoniaXX}, we have calculated the Casmir interaction between two rigid ``ct" rods at any separation (fig.~\ref{f.discrete}). Although the microscopic cutoff wavelength is explicitly involved in our calculations, we obtain universal results that only depend on $d/L$. At large separations, we recover the $(L/d)^4$ power-law obtained by Golestanian et al.~\cite{Golestanian96pre}. At small separations, we recover our result~(\ref{hard}). As can be seen on fig.~\ref{f.discrete}, the Casimir interaction potential reaches $-k_\mathrm{B}T$ for $d/L\simeq0.2$ (while using the asymptotic form would yield the overestimated value $d/L\simeq2$). 
Thus, two rods approaching each other at separations less than half their length will effectively attract each other due to the Casimir effect.

In order to check the regularization procedure mentioned after Eq.~(\ref{block}), we have recalculated the interaction between two rigid rods of the ``$\mathrm{\bar c t}$" type by using zeta-function regularization~\cite{Elizalde_book}, which is rather straightforward for such rods. To obtain the Casimir energy $\mathcal{F}^{\mathrm{\bar c t}|\mathrm{\bar c t}}_C(d)$, we calculate the partition function $Z=\int \mathcal{D}h\,\exp[-\beta \mathcal{H}]$ of the membrane patch $\mathcal{P}=[0,d]\times[0,L]$ situated between the rods (the free energy of the rest of the membrane is independent of $d$). The functional integral in $Z$ runs over the functions $h(\vect{r})$ that comply with the boundary conditions imposed by the rods, i.e., $\mathbb{D}^\mathrm{\bar c t} h(x,y)=0$ for $x\in\{0,d\}$, and $y\in[0,L]$. Let us assume, in addition, that $h(x,y)=0$ on the rods (it will not change the result since we have seen previously that the relative tilt of the rods is effectively frozen). Finally, let us take periodic boundary conditions with period $L$ in the $y$ direction.
Then, $\mathcal{H}=\frac{1}{2}\kappa\int_\mathcal{P} d\bm{r}\,h\mathcal{O} h$, where $\mathcal{O}=\nabla^4$ is the biharmonic operator on $\mathcal{P}$ with the above-mentioned boundary conditions. Since $\mathcal{O}$ is positive definite Hermitian, we can write $Z\propto1/\sqrt{\det(\mathcal{O})}$. Thus, discarding $d$-independent terms, we obtain $\mathcal{F}^{\mathrm{\bar c t}|\mathrm{\bar c t}}_C(d)=\frac{1}{2}k_\mathrm{B}T\,\mathrm{Tr}\ln(\mathcal{O})$. The eigenvalues of $\mathcal{O}$ are 
\begin{align}
\lambda_{nm}\,&=\,\pi^4\left(\frac{n^2}{d^2}+\frac{4\,m^2}{L^2}\right)^2, \hspace{1ex}n\in\mathbb{N}^*,\hspace{1ex} m\in\mathbb{Z},
\end{align}
and the associated eigenfunctions are $\sin(n\pi x/d)\times\exp(im2\pi y/L)$. Hence, we obtain
\begin{align}
\label{tracelog}
\mathcal{F}^{\mathrm{\bar c t}|\mathrm{\bar c t}}_C(d)\,&=\,\frac{1}{2}k_\mathrm{B}T\sum_{n=1}^\infty
\int_{-\infty}^\infty\!\upd m\,
\ln(\lambda_{nm})\,.
\end{align} 
While Eq.~(\ref{tracelog}) is divergent, its finite Casimir part can be extracted by calculating $\zeta_\mathcal{O}(s)=\sum_{n=1}^\infty\int_{-\infty}^\infty\!dm\,\lambda_{nm}^{-s}$, which is finite for $s>\frac{1}{4}$, and yields by analytical continuation:
\begin{equation}
\mathcal{F}^{\mathrm{\bar c t}|\mathrm{\bar c t}}_C(d)=-\frac{1}{2}k_\mathrm{B}T\lim_{s\to0}\zeta_\mathcal{O}'(s)=-k_\mathrm{B}T\frac{\pi}{12}\,\frac{L}{d}\,.
\end{equation}
This result is in agreement with the value $A^{\mathrm{\bar c t}|\mathrm{\bar c t}}=-\pi/12$ obtained above with the regularization of Li and Kardar~\cite{Li92}. This agreement also implies that a half-plane imposing the same boundary conditions as a rod will yield the same Casimir interaction (in the regime $L\gg d$). Physically, this equivalence is due to the fact that the Casimir interaction arises from the confinement of the fluctuation modes between the rods (or the half-planes), i.e., in the patch $\mathcal{P}$.

Finally, we have checked that the undulation instability discussed in Ref.~\cite{Golestanian96epl} can be recovered from the Casimir interaction calculated in this paper. Let us consider the fundamental deformation mode:
\begin{equation}
\delta x=u(y)=U\sin(\pi y/L')\,,\quad y\in[0,L']\,,
\end{equation}
where $\delta x$ is the in-plane displacement of each rod in the direction of the other rod. Inextensibility implies $L'=L-\pi^2U^2/(4L)+\mathcal{O}(U^3)$. In the limit $d\ll L$, we may use the proximity-force (or Derjaguin) approximation~\cite{Derjaguin34} to calculate the total free energy of the two rods:
\begin{align}
\label{inst}
\frac{\mathcal{F}_\ab{def}}{k_\ab{B}T}&\simeq
\int_{0}^{L'}\!\upd y\left(\frac{A^{\alpha|\beta}}{d-2u}+\ell_pu''^2\right)-\frac{A^{\alpha|\beta}L}{d}+\mathcal{O}(U^3)\nonumber\\
&=\frac{A^{\alpha|\beta}L}{d}\,\frac{4U}{\pi d}\left(1+\frac{\pi U}{2d}\right)+\frac{\pi^4\ell_p}{2L}\,\frac{U^2}{L^2}+\mathcal{O}(U^3).
\end{align}
Minimizing $\mathcal{F}_\ab{def}$ with respect to $U$, we find that the equilibrium deformation amplitude $U_\ab{eq}$ reaches $d/2$, i.e., that the two rods touch each other when $d<d'_c$, where
\begin{equation}
\frac{d_c'}{L}=\left(\frac{-4(2+\pi)A^{\alpha|\beta}}{\pi^5}\right)^{1/3}\left(\frac{L}{\ell_p}\right)^{1/3}.
\end{equation}
We thus obtain an instability threshold very similar to that of eq.~(\ref{dc}). For instance, for two ``ct" or ``$\mathrm{\bar c\bar t}$" rods, with $A^{\alpha|\beta}\simeq-0.46$, the prefactor of $(L/\ell_p)^{1/3}$ is $0.31$, while in eq.~(\ref{dc}) it is $0.27$.

In this paper, we have assumed that the effect of the membrane tension $\sigma$ was negligible with respect to that of its bending rigidity $\kappa$. This hypothesis is valid in usual flaccid membranes ($\sigma\approx10^{-7}\,\mathrm{N/m}$) for rods shorter than $(\kappa/\sigma)^{1/2}\approx 1\,\mu\mathrm{m}$. However, $\sigma$ can be increased up to about $10^{-3}\,\mathrm{N/m}$ by applying an external tension to the membrane: in this case, the effect of tension will become important. The methods presented here in the bending-rigidity--dominated regime can also be applied to the tension-dominated limit, where $L$ and $d$ are much larger than $(\kappa/\sigma)^{1/2}$. For instance, the interaction between two ``ct'' rods imposing $h_{yy}=0$ is then still given by Eq.~\ref{F2}, but with $A^\mathrm{ct|ct}=-\pi/24$. This result can be obtained both through the method based on Ref.~\cite{Li92} and through zeta-function regularization. This tension-dominated interaction is relevant in liquid-vapor interfaces~\cite{Lehle07}. A general calculation of the Casimir-like interaction between two rods, including both $\kappa$ and $\sigma$, is beyond the scope of this paper. This problem has been addressed very recently in the simpler case of two small circular inclusions~\cite{Lin}.

In summary, we have studied the Casimir interaction between two parallel rods adsorbed on a fluid membrane characterized by its bending energy and subject to thermal fluctuations. We have considered four types of boundary conditions, depending on whether the membrane can twist along the rod and/or curve across it. Whatever their type, rigid rods of length $L$ at short separations $d\ll L$, experience an attractive Casimir interaction scaling as $k_\ab{B}TL/d$. In this regime, two of the four types of rods are equivalent, which yields six universal Casimir amplitudes. When taking into account the finite persistence length $\ell_p$ of the rod, we find, in agreement with Ref.~\cite{Golestanian96epl}, that at separations less than $d_c\sim L(L/\ell_p)^{1/3}$ two rods with fixed ends will bend toward each other and finally come into contact because of the Casimir interaction. 
Besides, we have shown that rods with a very soft out-of-plane bending rigidity, but a large in-plane bending rigidity, still experience a power-law Casimir interaction (albeit small), which scales as $d^{-3}$. For a large, but not infinite out-of-plane bending rigidity, an effective power law in $d^{-\alpha}$ with $\alpha$ close to $1$ would be observed. This nuances the conclusion of Ref.~\cite{Golestanian96epl} according to which a finite out-of-plane rigidity, no matter how large, should destroy the long-range Casimir interaction. 
We have also calculated numerically the Casimir interaction between two rods at any separations, finding at $d/L\approx0.5$ a crossover between the short separation $d^{-1}$ behavior and the large separation $d^{-4}$ behavior. We have shown, in the case of the simplest boundary conditions, that at separations $d/L\io0.2$ the Casimir attraction reaches $k_\ab{B}T$. In conclusion, due to the very long range of the Casimir interaction, long rods adsorbed on an effectively tensionless membrane will actually attract each other---by the Casimir effect---when their separation becomes significantly smaller than their length.

\end{document}